\documentclass[prl, twocolumn, superscriptaddress,showpacs,amsmath,amssymb,floatfix]{revtex4}
\usepackage{graphicx,color}
\usepackage{epsfig}
\usepackage{bm}
\newcommand{\scl}{S_{\mbox{\rm\tiny cl}}}

\begin{document}

\title{Classical Dynamics of Quantum Entanglement}

\author{Giulio Casati}
\affiliation{CNISM, CNR-INFM \& Center for Nonlinear and Complex Systems,
Universit\`a degli Studi dell'Insubria, Via Valleggio 11, 22100 Como, Italy}
\affiliation{Istituto Nazionale di Fisica Nucleare, Sezione di Milano,
via Celoria 16, I-20133 Milano, Italy}
\author{Italo Guarneri}
\affiliation{ Center for Nonlinear and Complex Systems,
Universit\`a degli Studi dell'Insubria, Via Valleggio 11, 22100 Como, Italy}
\affiliation{Istituto Nazionale di Fisica Nucleare, Sezione di Pavia,
via Bassi 6, I-27100 Pavia, Italy}
\author{Jose Reslen}
\affiliation{Department of Physics, National University of Singapore, Singapore 117543}

\date{\today}


\begin{abstract}

We numerically analyze the dynamical generation of quantum entanglement in a system of 2 interacting particles, started in a  coherent separable state, for decreasing values of $\hbar$. As $\hbar\to 0$  the  entanglement entropy, computed at any finite time,  converges to a  finite nonzero value. The limit law that rules the time dependence of entropy is well reproduced
by purely classical computations. Its general features  may then be explained by simple classical arguments, which expose
the different ways entanglement is generated in systems which are classically chaotic or regular.

\end{abstract}
\maketitle

The development of quantum chaology has shown that quantum dynamics is deeply influenced by classical chaotic motion.
  It was then natural to inquire, how is the dynamical generation  of quantum entanglement affected by the chaotic or regular nature of the underlying classical dynamics  \cite{pellegrino,lista,Novaes,Gong,PJ04,ZnPr05,Nature,Miller};
 an especially intriguing aspect of this question  being that quantum entanglement does not have a classical counterpart.
Since the early paper by Furuya {\it et al.} \cite{pellegrino}, most investigations about this subject have supported the view that regular classical dynamics leads to slower dynamical generation of entanglement (DGE)\cite{lista}. The opposite point of view, that DGE is scarcely affected by the nature of classical dynamics, has also been surmised \cite{Novaes}; nevertheless     there are convincing reasons to believe that the way quantum entanglement  is dynamically generated does indeed have a classical counterpart\cite{Gong,AngFu05,Matz}. Notably, DGE may be studied by resorting to a phase-space formulation of quantum mechanics \cite{Gong}; and then  the purity of reduced states, which is a measure of (dis-)entanglement, can be directly obtained  from the phase-space functions which represent such quantum states in the classical phase spaces. Letting such  functions evolve according to the classical Liouville dynamics a classical counterpart to DGE is obtained\cite{Gong}.
In a similar line of thought,  quasi-classical analytical methods  have been implemented to investigate the decay of purity of  reduced states in integrable systems\cite{PJ04,ZnPr05}.\\
In this Letter we study the $\hbar\to 0$ limit of DGE.  We thus demonstrate that DGE has a direct classical analog, in the following sense. We start a system  of two interacting particles  in a separable coherent state, and  numerically compute the growth in time $t$ of their entanglement ,  as measured by the von Neumann entropy $S(t,\hbar)$ of their  reduced states. In the limit $\hbar\to 0$, the evolving quantum state tends to a  classical pure state ({\it i.e.} a Dirac $\delta$-function in phase space,  supported in a point of the classical trajectory), and yet its entanglement entropy does not tend to
zero\cite{alr}; instead it converges to a finite value, and we find that this value can be  approximated using    Hamilton's equations.

We consider a  Toda model of two interacting particles  moving in a line under
the following Hamiltonian:
\begin{equation}
\label{todham}
{H}= \frac{{p}_1^2}{2 m_1} + \frac{{p}_2^2}{2 m_2} + e^{-{q}_1} + e^{-\left({q}_2-{q}_1 \right)} +  e^{{q}_2} - 3.
\end{equation}
Relevant parameters are the energy $E$,  and the mass ratio $m_1/m_2$. For $m_1=m_2$ the model is completely integrable no matter how large is the energy (Fig.1a), while for any other mass ratio (except limit values $0,\infty$)
the dynamics exhibits a transition from integrability to chaos as the energy E is increased \cite{Toda}. In particular for $m_2/m_1 =0.54$ motion is predominantly chaotic already at energy $E= 7$ (Fig.1b).

  To compute the quantum evolution  we have first analytically calculated  matrix elements of the quantum Hamiltonian (which is obtained from (\ref{todham}) by canonical quantization) over the energy eigenbasis of two uncoupled harmonic oscillators of unit mass, and then we have diagonalized  the matrix thus obtained. With the chosen  matrix size, the
computed eigenbasis  was verified  to support the chosen initial states within negligible errors.
 Such initial states were  separable coherent states, centered at $q_1=q_2=0$ and $p_1=p_2=\sqrt{E}$ in the regular case ($m_1=m_2=1$), and $q_1=q_2=0$, $p_1=\sqrt{E}$, $p_2=-\sqrt{m_2 E}$ in the chaotic case ($m_1=1, m_2=0.54$). The latter initial condition is inside the chaotic sea.
As time goes on the particles get entangled, due to interactions. A measure of their entanglement at time $t$ is given by the von Neumann entropy $S(t) = -{\text {Tr}}(\hat{\rho_1}\ln {\hat{\rho_1}}) $ where $\hat{\rho_1}$ is the reduced density matrix of particle 1.
The this way computed time dependence of entanglement is shown in Fig. \ref{fig:2} for three different values of $\hbar$.
 A  first observation is that integrable and chaotic cases can hardly be told from each other by just comparing
 the degrees of entanglement to which they give rise. In this respect it is worth noting that the dependence of entanglement on initial conditions, which was  observed in several papers ({\it e.g.} \cite{Nature}) as the initial state was moved from the chaotic sea to an integrable island, is \emph{not} a signature of chaos. Instead it is merely due to a decrease of the number of accessible quantum states,  as one approaches the periodic orbit in the center of the island. Indeed, a similar, significant  decrease of entanglement  is recorded even in the completely integrable case of our model (not shown),  starting with initial conditions that approach  the central periodic orbit of the stable island; {\it e.g.}, along a vertical line ($q_1=1.2$) in Fig.1a.
\begin{figure}
\begin{center}
\includegraphics[width=0.3\textwidth,angle=-0]{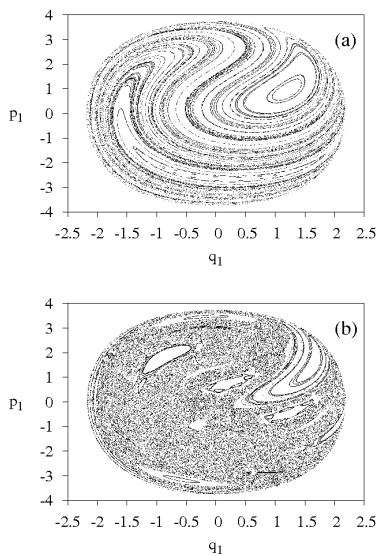}
\end{center}
\caption{Surfaces of section for the Toda model at energy $E=7$. (a) integrable case with  masses $m_1=m_2=1$; (b) mixed phase space for $m_1=1,m_2=0.54$.}
\label{fig:1}
\end{figure}

  However, the most interesting observations are:  first, as $\hbar$ decreases,   the curves  $S(t,\hbar)$, which describe the entanglement of the quantum Toda system as a function of time $t$ at a given value of $\hbar$,  come closer and closer to a nontrivial limit curve (Fig. \ref{fig:2}, full curve). Second,  this curve is well approximated by a purely classical construction, to be described next.

From the classical Gaussian ensemble in the 4-dimensional phase space that corresponds to the initial quantum coherent state we have generated a sample of $M$ initial points. To compute the evolution of this initial ensemble we have integrated the classical Hamilton equation,  using a fourth-order Runge-Kutta routine. The $4$-dimensional ensembles thus computed at times $t>0$
were projected onto
 "reduced"  ensembles in
the $(q_1,p_1)$ plane. To obtain a classical analog for the Von Neumann entropy of quantum reduced states, the
 Boltzmann-Shannon
entropies $\scl (\delta,t)$ (with $\delta = \hbar$) were computed, using a fixed partition of
the $2$-dimensional phase space in square cells  of side $\sqrt{\delta}$:
\begin{equation}
\scl(\delta,t) = -\sum_i \frac{w_i(t)}{M} \ln {\frac{w_i(t)}{M}},
\label{eq:16}
\end{equation}
where $w_i(t)$ is the number of points in the reduced ensemble , which were found  in the $i$'th cell at time $t$. We
used enough cells to guarantee hosting of all points at all times. This classical entropy is expected to somehow approximate the entropy of the quantum reduced ensemble, provided the latter ensemble mixes a sufficiently large number of states.

  We numerically observe that, as $\delta\to 0$, the curve $\scl(\delta,t)$ approaches a limit curve $S_{\infty}(t)$. This seemingly crude classical construction works surprisingly well; indeed,  the limit classical  curve is more and more closely approached by the curves of quantum entanglement as  $\hbar\to 0$. The approximation is  so good,  that even  finer details of the quantum curves, such as  superimposed almost regular oscillations (fig.2),  can be explained in terms of classical phase space structures, as we shall presently explain.

Before coming to that, it is important to stress that our findings do not involve any contradiction. Measures of quantum entanglement need not tend to $0$ in the  $\hbar\to 0$ limit,  as shown by simple examples: {\it e.g.}, a system of two identical harmonic oscillators, in the pure state
$\tfrac1{\sqrt 2}H_0\otimes H_0+\tfrac1{\sqrt 2}H_1\otimes H_1$, where $H_0, H_1$ are the ground state and the 1st excited state of a harmonic oscillator, has entanglement entropy equal to $\ln(2)$, independently of the value of $\hbar>0$, even though in the limit $\hbar\to 0$  its Wigner function tends to a Dirac $\delta$ supported in the origin of the 4-dimensional phase space. On the other hand,  the meaning of the above described classical construction is an interesting question in purely classical terms, that may be related to a recently proposed classical separability entropy \cite{Proz11}.

 In Figs. 3,4 we show the numerically computed classical and quantum entropies  respectively.
\begin{figure}
\begin{center}
\includegraphics[width=0.4\textwidth,angle=0]{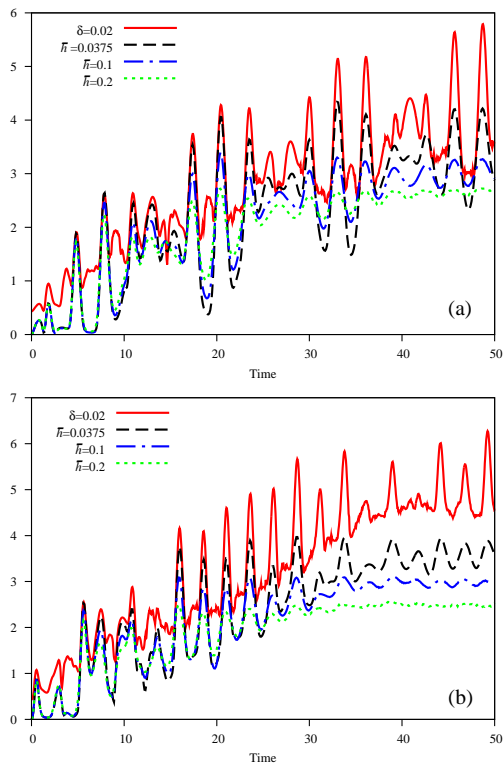}
\end{center}
\caption{Entanglement entropy vs time for $E=7$ and decreasing values of $\hbar$ in a regular (a) and in a chaotic (b) case. The full curve is  classical entropy.
}
\label{fig:2}
\end{figure}
\begin{figure}
\includegraphics[width=0.4\textwidth,angle=0]{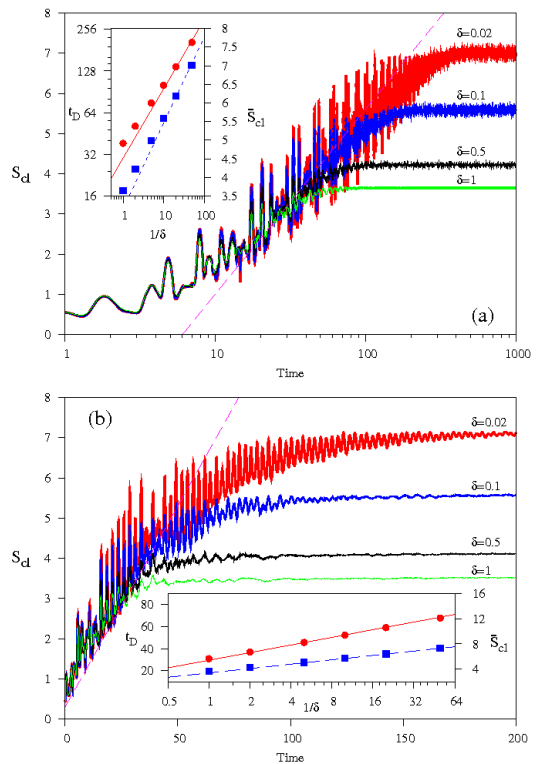}
\caption{Classical entropy $S_{cl}(t)$ vs time for different values of the cell area $\delta$ for $E=7$. Above, an integrable case; the dashed line is given by $const + 2 \ln(t)$. Below, a chaotic case; the dashed line is given by $const +  t/10$.
Insets: saturation time $t_D$ (circles) and saturation value $\overline{S_{cl}}$ (squares) versus $\delta$. In the regular case, $t_D\approx const + 1/\sqrt{\delta}$, $\overline{S_{cl}}\approx const + \ln(1/\delta)$. In the chaotic case:  $t_D \approx const +10 \ln(1/\delta)$;  $\overline{S_{cl}}\approx const + \ln(1/\delta)$.}
\label{fig:3}
\end{figure}
\begin{figure}
\includegraphics[width=0.4\textwidth,angle=0]{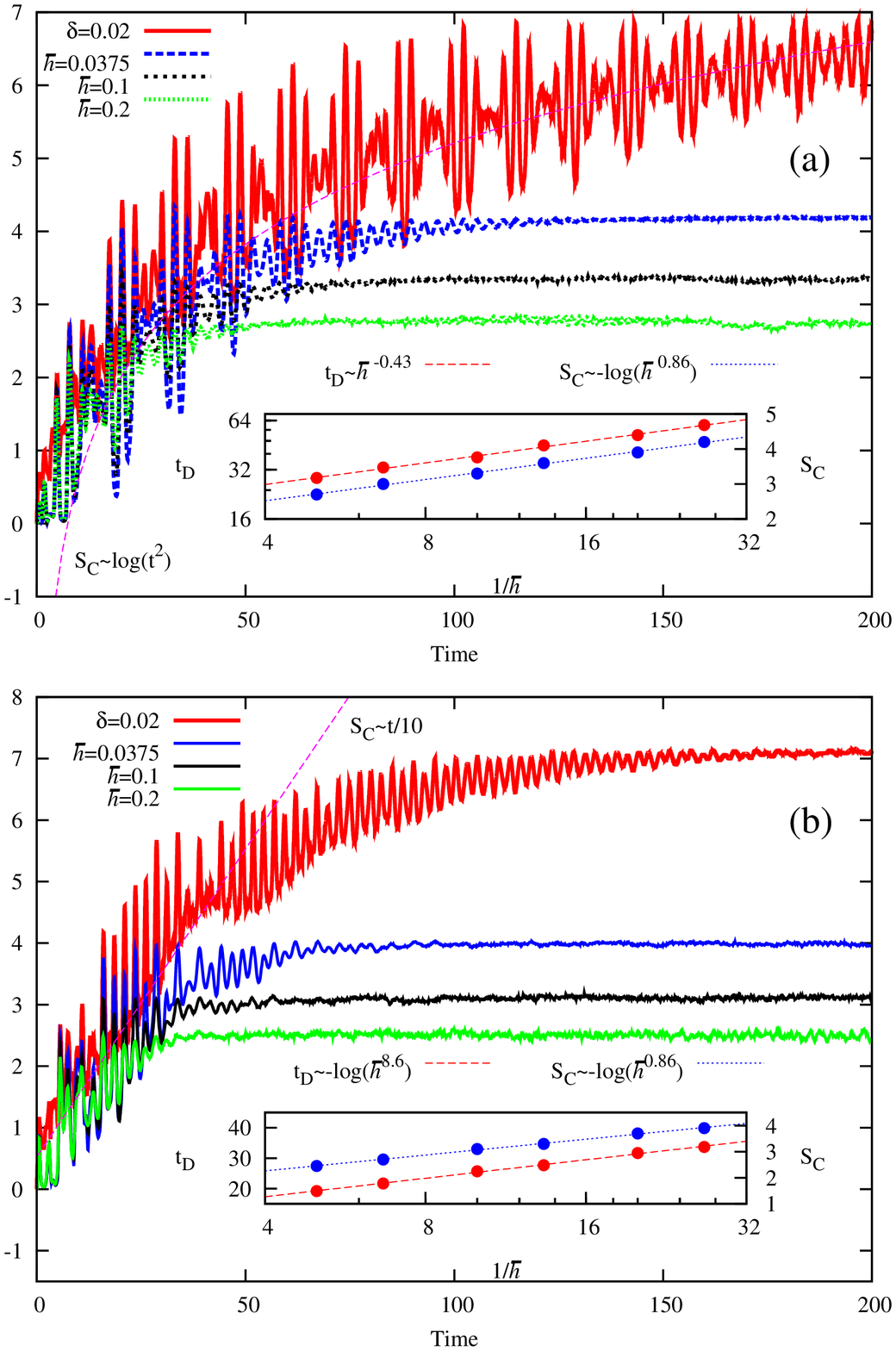}
\caption{Entanglement entropy $S(t)$ vs time for different values of $\hbar$, in an integrable case (above) and in a  chaotic case (below). The continuous curves are the classical entropy $S_{cl}$ for $\delta=0.02$ as in fig 4.
Insets: saturation time $t_D$ (circles) and saturation value $\overline{S}$ (squares) of  entanglement, versus $\hbar$. In the regular case: $t_D\approx const + 1/\sqrt{\hbar}$, $\overline{S}\approx const + \ln(1/\hbar)$. In the chaotic case:  $t_D \approx const +10 \ln(1/\hbar)$;  $\overline{S}\approx const + \ln(1/\hbar)$}
\label{fig:4}
\end{figure}
All the shown curves share some general qualitative features: notably, at long times they approach  saturation values,  via an average monotonic trend with strong  oscillations superimposed. Such oscillations look surprisingly regular even in chaotic cases,  and gradually decrease in time; in the classical cases, at saturation they are on the order of fluctuations due to finite statistics. As the cell area $\delta$ is decreased,
for times less than  a time scale $t_D(\delta)$ a
limit curve is approached, which is different in the regular and in the chaotic cases.  The saturation time $t_D$ may  be thought of as the time when a finite $\delta$ curve
departs from the limit behaviour.  For times $t>t_D(\delta)$,  entropy curves approach a saturation value,  which scales proportional to $1/\sqrt{\delta}$ in the regular case and to $\ln(1/\delta)$ in the chaotic case (see insets  in Fig.3). To infer estimates of saturation times from numerical data, $t_D(\delta)$ was defined as the time when the average growth (represented by dashed lines) of the limit curve reaches the saturation value at the given $\delta$.\\
Such numerical observations are easily explained by interpreting  entropy as the logarithm of an effective number of  populated cells.
To this end, one may first replace  the initial Gaussian ensemble by a single $4$-dimensional cell of size $\sqrt\delta$. Saturation is then an obvious consequence of the fact that only a finite number of cells are available, due to energy conservation and compactness of the energy shell. The saturation value of entropy is  expected to scale with $\delta$  like the logarithm of that number, {\it i.e.}, in the leading order, like $C\ln(1/\delta)$ with $C=1$,  both in regular, and in chaotic cases. This is consistent with numerically observed values $C\simeq 0.86$ and $C=0.93$ for the regular and the chaotic case respectively.\\
The average growth of entropy at large times, yet still far from saturation,  may be understood similarly.
The initial cell intersects different $3$-dimensional manifolds of constant energy. In the chaotic case, evolution stretches  the $3$-dimensional section at energy $E$ along the unstable direction and contracts it
 along a stable one , at an exponential rate given by the maximal Lyapunov exponent $\lambda_E$. Hence the number of cells which are populated by the projection of this section grows roughly like $\exp(\lambda_E t)$.
 The number of cells populated by  the projections of all sections grows like a continuous superposition of exponentials with rates $\lambda_E$ varying in a small interval of size $\propto\sqrt{\delta}$, so, in the leading order,  entropy grows  proportional to $t$ times the average Lyapunov exponent, independently of $\delta$.  Equating this to the saturation value
   the order of magnitude estimate $t_D\sim \lambda^{-1}\ln(1/\delta)$ is obtained, consistently with  numerical results.

In the regular case, the initial $4$-dimensional cell intersects several invariant $2$-tori. Each  $2$-dim section glides uniformly in time along the  torus where it belongs, carried by the linear flow which is associated with  the frequency vector of that torus; so its projection on the single particle phase space is just deformed quasi-periodically in time,  with no average growth of  the number of populated 2-dim cells.
The number of $2$-dimensional cells populated by all such projections nevertheless increases, because different sections move with different frequency vectors due to nonlinearity; so, although all sections were originally piled up over the same 2-dim cell,
their projections slide apart linearly in time, $\propto t\sqrt{\delta}$ (because the mismatch
of frequency vectors is on the order of the linear size of the cell). As sections are a $2$-parameters family , the $2$-dimensional area covered by such projections increases quadratically in time :
$\sim$const.$\times \delta t^2$ in the leading order.
Hence, the number of occupied cells of area $\delta$ increases $\sim$const.$\times t^2$, and entropy grows , in the leading order, like $2\ln(t)$,
independently of $\delta$. At time $t_D$ this should be on the order of magnitude of the saturation value $\ln(1/{\delta})$, whence the observed scaling of the saturation time $t_D\sim1/\sqrt{\delta}$ follows .

As shown in Fig.4, the behavior of quantum entanglement is quite similar. The same scaling  with $\hbar$ of  deviation times and  saturation values emerges as  the classical limiting curves are approached, both in the integrable and in the chaotic case \cite{Miller}.

The oscillatory pattern which is seen in all figures, superimposed on the just discussed average behavior, is qualitatively explained as follows. In the regular case: the projection of each invariant torus on the $(q_1,p_1)$ plane has caustics at its boundaries, because there the  tangent plane of the torus is "vertical" . As sections of the initial ensemble slide around  their tori, the areas of their  projections change quasi-periodically in time, and in particular become quite small  anytime they come close to caustics. So long as different sections of the initial ensemble have not yet strongly dephased from one another, this produces  a regular sequence of minima in the number of occupied cells.
In the chaotic case: observation of the projected ensembles in the $(q_1,p_1)$ plane, at  times which correspond to minima in the regular sequence of oscillations, reveals that phase-space points are then concentrated near a caustic-like line, that betrays presence of remnants of a broken torus. The projected ensemble appears to quasi-periodically concentrate near this line, until  the $4$-dimensional ensemble eventually leaks  through the broken torus to invade most of the energy shell.\\
We finally note that, while the quantum reduced entropy is the same for both particles, in general our classical reduced entropies approach each other only in the limit $\delta\to 0$.\\
In conclusion, we have shown that there is a close similarity between the laws which rule the dynamical generation of quantum and classical reduced entropies. The classical  laws admit of simple heuristic explanations, which provide clues to  understanding  the quantum ones. Such arguments  predict  differences between the ways entanglement is generated in classically regular  and chaotic regimes, that  should be observable for generic quantum multi-partite systems in the quasi-classical regime.


GC and IG acknowledge support by
MIUR-PRIN 2008 and by Regione Lombardia. JR acknowledges
support from  grant projects (R-144-000-276-112) and
(R-710-000-016-271).


\begin{thebibliography}{0}
\bibitem{alr}{A similar behavior has been observed in different model systems in refs.
\cite{AngFu05},\cite{Matz}.}
\end{thebibliography}
\end{document}